\documentclass[prb,twocolumn]{revtex4}
\usepackage{graphicx}
\usepackage{hyperref}
\usepackage{amssymb}
\usepackage{caption}
\usepackage{dcolumn}
\usepackage{float}
\usepackage{bm}
\usepackage{relsize}
\usepackage{booktabs}
\usepackage{amsmath} 
\usepackage[utf8]{inputenc}
\usepackage{lmodern}
\usepackage{color}

\newcommand{\nn}{\nonumber}

\newcommand{\e}{\mathrm{e}}

\DeclareMathAlphabet{\bi}{OML}{cmm}{b}{it}

\def\be{\begin{equation}}
\def\ee{\end{equation}}
\def\bearr{\begin{eqnarray}}
\def\eearr{\end{eqnarray}}

\usepackage{tikz}

\begin{document}
\title{Photoinduced valley and electron-hole symmetry breaking in 
$ \alpha$-$T_3$ lattice: The role of a variable Berry phase}
\bigskip
\author{Bashab Dey and Tarun Kanti Ghosh\\
\normalsize
Department of Physics, Indian Institute of Technology-Kanpur,
Kanpur-208 016, India}
\begin{abstract}
We consider $\alpha$-$T_3$ lattice illuminated by intense circularly 
polarized radiation in terahertz regime. We present quasienergy band 
structure, time-averaged energy spectrum and time-averaged density of 
states of $\alpha$-$T_3$ lattice by solving the Floquet Hamiltonian numerically. 
We obtain exact analytical expressions of the quasienergies at the 
Dirac points for all values of $\alpha$ and field strength. We find that 
the quasienergy band gaps at the Dirac point decrease with increase of
$\alpha$. Approximate forms of quasienergy and band gaps at single and 
multi-photon resonant points are derived using rotating wave approximation.
The expressions reveal a stark dependence of quasienergy on the Berry phase 
of the charge carrier. 
The quasi energy flat band remains unaltered in 
presence of radiation for dice lattice ($\alpha=1$). However, it acquires a 
dispersion in and around the Dirac and even-photon resonant points when 
$0<\alpha<1$. The valley degeneracy and electron-hole symmetry in the 
quasienergy spectrum are broken for $0<\alpha<1$. 
Unlike graphene, the mean energy follows closely the linear dispersion of 
the Dirac cones till near the single-photon resonant point in dice lattice.
There are additional peaks in the time-averaged density of states at the Dirac point
for $0 < \alpha \leq 1$.
\end{abstract}

\maketitle

\section{Introduction}
In recent years, dynamical effect of an intense AC field on
electronic, transport and optical properties in quantum 
two-dimensional materials having Dirac-like spectrum has drawn much interest 
\cite{Hanggi,Eckardt,Efetov,Efetov1,Lopez,Oka,Oka1,Zhao,Kibis,Wu,Schilman,Gupta}. 
It is seen that intense time-periodic field substantially changes
the energy band structure by photon-dressing and consequently the
topological properties of materials.
Inducing gap in Dirac materials is an important issue for
electronic devices.    
A stationary energy gap appears at the Dirac points
under a circularly polarized radiation \cite{Oka,Zhao,Kibis}. 
Also, the gaps appear in the quasienergy spectrum \cite{Wu} due 
to single-photon and multi-photon resonances, which decreases with
increase in momentum.
Oka and Aoki showed that photovoltaic Hall effect can be induced in graphene under 
intense ac field \cite{Oka}, even in absence of
uniform magnetic field. The energy gap at the Dirac point closes 
as soon as the spin-orbit interaction in graphene monolayer is
taken into account \cite{Schilman}. The optical conductivity of graphene monolayer 
under intense field has been reported to show multi-step-like behavior due to
sideband modulated optical transitions \cite{Wu}.
A photo induced topological phase transition in silicene has been proposed
by Ezawa  \cite{Ezawa}.
The photoinduced zero-momentum pseudospin polarization, 
quasienergy band structure and time-averaged density of states (DOS) of the charge 
carriers in monolayer silicene have also been studied \cite{Schilman-silicene}.

\begin{figure}[htbp]
\includegraphics[trim={0cm 1cm 0cm  2cm},clip,width=8cm]{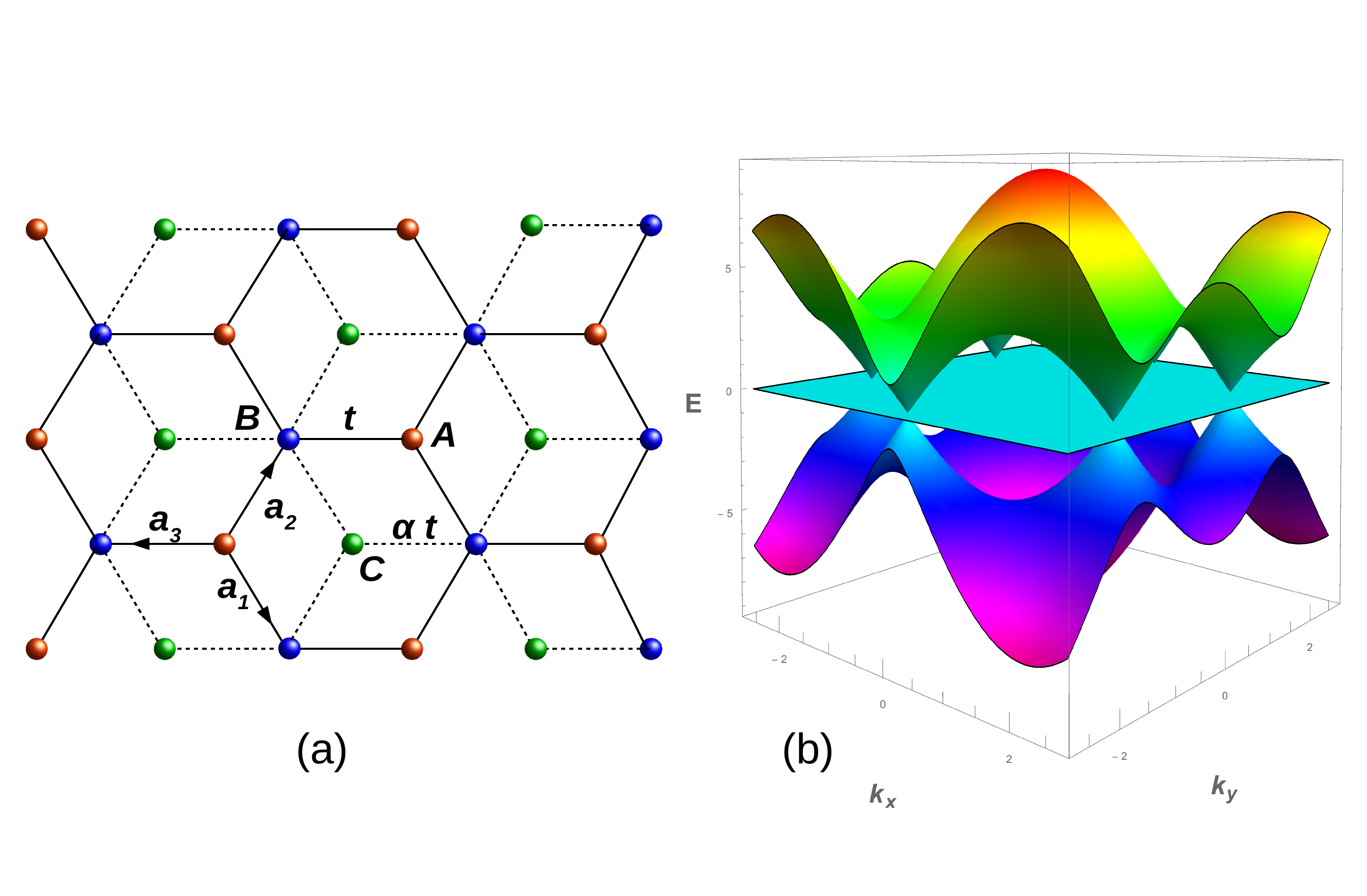}
\caption{(a) Sketch of the $\alpha$-$T_3$ lattice. (b) Band structure of 
$\alpha$-$T_3$ lattice using tight-binding lattice.}
\label{lattice}
\end{figure}

There exists an analogous lattice of graphene \cite{graphene}, known as 
$\alpha$-$T_3$ lattice, in which quasiparticles are described by the 
Dirac-Weyl equation. The $\alpha$-$T_3$ lattice, as shown in Fig.\ref{lattice}(a), 
is a honeycomb lattice with two sites (A,B) and an additional site (C) at the 
center of each hexagon. The C sites are bonded to the alternate corners of the 
hexagon, say B sites. The hopping parameter between A and B sites 
is $t$ and that between C and B sites is $\alpha t$. 
The sites in such a lattice can be subdivided into two categories on 
the basis of number of nearest neighbors$-$\textit{\textbf{hub}} (B) sites with 
coordination number 6 and \textit{\textbf{rim}} (A,C) sites with coordination number 3. 
The rim sites form hexagonal lattice with no bonds among them. The hub sites form 
a triangular lattice. Each hub site is connected to 6 rim sites out of which 3 are 
equivalent. The hopping parameter alternates between $t$ and $\alpha t$ among the 6 
hub-rim bonds from a single hub site. The $\alpha=0$ results in the honeycomb lattice 
resembling monolayer graphene, which corresponds to Dirac-Weyl system with 
pseudospin-1/2. On the other hand, $\alpha = 1$ leads to the well-studied $T_3$ or 
dice lattice with pseudospin-1 \cite{Sutherland,Vidal,Korshunov,Rizzi,Urban,JDMalcolm,Wolf,Vigh}. 
Tuning of $\alpha$ from 0 to 1 gradually allows us 
to study the continuous changes in the electronic properties of massless fermions.

The dice lattice can naturally be built by growing trilayers
of cubic lattices (e.g. SrTiO${}_3$/SrIrO${}_3$/SrTiO${}_3$) in (111) 
direction \cite{Ran}. 
An optical dice lattice can be produced by a suitable arrangement 
of three counter-propagating pairs of laser beams \cite{Rizzi}. 
The $\alpha$-$T_3$ optical lattice can be realized by dephasing one of the pairs 
of laser beams with respect to other two \cite{Raoux,Rizzi}.
The Hamiltonian of Hg${}_{1-x}$Cd${}_x$Te quantum well
can also be mapped to that of low-energy $\alpha$-T${}_3$ model with 
effective $\alpha=1/\sqrt{3}$ on appropriate doping \cite{Malcolm}.

Recently, a list of physical quantities like 
orbital susceptibility \cite{Raoux}, optical conductivity \cite{Illes,Illes,Cserti}, 
magnetotransport properties \cite{Tutul,Malcolm,Duan,Firoz}, Klein
tunneling \cite{Urban,Klein} and wave-packet dynamics \cite{Tutul1} in 
$\alpha$-$T_3$ lattice have been studied extensively.  
The Berry phase has become indispensable ingredient in modern condensed matter 
physics due to its strong influence on magnetic, transport and optical properties 
\cite{Niu-RMP}. For example, the variation of the orbital susceptibility 
with $\alpha$ is a direct consequence of the variable Berry phase of the 
$\alpha$-T${}_3$ lattice \cite{Raoux}. It has been pointed out that the 
quantization of the Hall plataues \cite{Tutul,Malcolm,Duan} and behavior of 
the SdH oscillation \cite{Tutul} change with the Berry phase of the 
$\alpha$-T${}_3$ lattice. The Berry phase dependence of the longitudinal optical 
conductivity of the $\alpha$-T${}_3$ lattice has also been reported \cite{Illes}.

In this work, we study quasienergy band structure, time-averaged energy 
spectrum and time-averaged density of states of $\alpha$-$T_3$ lattice 
irradiated by circularly polarized light. We provide exact and approximate 
analytical expressions of the quasienergies at the Dirac points as well as 
at resonant $k$ points for all values of $\alpha$, respectively. The valley 
degeneracy and the electron-hole symmetry are destroyed by the circularly 
polarized radiation for $0<\alpha<1$. We establish a direct connection between 
the quasienergy spectrum and the variable Berry phase, which is responsible 
for the broken valley degeneracy. The quasienergy gap at the Dirac point 
decreases with $\alpha$. The behavior of the time-averaged energy and 
time-averaged density states for $ 0 < \alpha \leq 1 $ are appreciably different 
from that of monolayer graphene.

This paper is organized as follows. In section II, we present preliminary 
information of the $\alpha$-$T_3$ lattice.
In section III, we solve Floquet eigensystem for $\alpha$-$T_3$ lattice driven
by circularly polarized light.
In particular, we present numerical and analytical results of quasienergy bands
and the corresponding band gaps.
In section IV,  the results of time-averaged energy spectrum and time-averaged 
density of states are presented.
In section V, we discuss main results of our study.

\section{Basic information of $\alpha-T_3$ lattice}
The rescaled tight-binding Hamiltonian of the system considering only nearest 
neighbour (NN) hopping is given by 
\begin{equation}
H_0(\textbf{k})=\left(\begin{array}
{ccc}
0 & t f^*(\textbf{k}) \cos\phi & 0
\\ t f(\textbf{k}) \cos\phi & 0 & t f^*(\textbf{k}) \sin\phi
\\0 & t f(\textbf{k}) \sin\phi & 0
\end{array}\right)
\end{equation}
where $t$ is the NN hopping amplitude, $\alpha$  is parameterized by the
angle $\phi$ as $\alpha=\tan\phi$ and
$f(\textbf{k}) = \sum \limits_{j=1}^{3} \e^{i \bf{k} \cdot {\bf a}_j}$.
Here, $\bf{a}_j$'s are the position vectors of the three nearest neighbors 
with respect to the rim site. 
Diagonalising the Hamiltonian gives three energy bands ($E_{\lambda}$) 
independent of $\alpha$ \cite{eucledian}: $E_{\pm}(\textbf{k}) = \pm t |f(\textbf{k})|$ and 
$E_{0}(\textbf{k})=0$. Here $ \lambda = +1,0,-1$ correspond to the conduction,
flat and valence bands, respectively.
A unique feature of its band structure is that a flat band $E_{0}({\bf k})$ is 
sandwiched between two dispersive bands $E_{\pm}({\bf k})$ which have 
electron-hole symmetry. 
The nondispersive band also appears in the Lieb \cite{Dagotto,Shen,Apaja,Goldman} as well
as Kagome models \cite{Green}.
Recently, the dispersionless flat band has been engineered in a photonic Lieb lattice 
formed by a two-dimensional array of optical waveguides \cite{Lieb-exp,Lieb-exp1}.
The flat band remains dispersion-less for all values of $\alpha$ and ${\bf k}$.
On the other hand, the dispersion of the conduction and valence bands is 
identical to that of graphene. The full band structure is shown in 
Fig. \ref{lattice}(b).

The low-energy Hamiltonian around the two inequivalent Dirac points
${\bf K} $ and $ {\bf K}^{\prime}$
can be written as
\be
H_0^{\mu}({\bf k}) = \hbar v_f {\bf S}(\alpha) \cdot {\bf k},
\ee
where $v_f = 3 a t/(2 \hbar) $, ${\bf k} = \mu k_x \hat {\bf x} + k_y \hat {\bf y} $ with 
$\mu = \pm 1$ refers to the $\textbf{K}$ and $\textbf{K}^{\prime}$ valleys, respectively 
and the components of the spin matrix ${\bf S}(\alpha)$ are defined as
\begin{equation}
S_x(\alpha) = \left(\begin{array}
{ccc}
0 & \cos\phi & 0 \\  
\cos\phi & 0 & \sin\phi \\ 
0 & \sin\phi & 0
\end{array}\right),
\end{equation}
\begin{equation}
S_y(\alpha) =  \left(\begin{array}
{ccc}
0 & -i \cos\phi & 0 \\
i \cos\phi & 0 & -i \sin\phi \\ 
0 & i \sin\phi & 0
\end{array}\right).
\end{equation}

In the vicinity of the two Dirac points,
$E_{\pm}(\textbf{k})$ are linear in ${\bf k}$ i.e.
$E_{\pm}(\textbf{k}) = \pm \hbar v_f |\textbf{k}|$,
implying massless excitations around the Dirac points,
as in the case of graphene.

In contrast to the band structure, the normalized eigen vectors 
$ \psi_{\textbf{k},\lambda} $ depend on $\alpha$ and are given by 
\begin{equation}
\psi_{\textbf{k},\pm}=\frac{1}{\sqrt{2}}\left(\begin{array}
{ccc}
\mu \cos\phi e^{-i \mu \theta_\textbf{k}}\\
\pm 1 \\
\mu \sin\phi e^{i \mu \theta_\textbf{k}}\end{array}\right), 
\psi_{\textbf{k},0}=\left(\begin{array}
{ccc}
\sin\phi e^{-i \mu \theta_\textbf{k}}\\
0\\
-\cos\phi e^{i \mu \theta_\textbf{k}}\end{array}\right), \nn
\end{equation}
where  $\theta_{\bf k} = \tan^{-1}(k_y/k_x)$. Moreover, 
the elements of the spinors from top to bottom represent the 
probability amplitude of staying in sublattices A (rim), B (hub) and C (rim),
respectively. The flat band wavefunction exhibits that the 
probability amplitude of an electronic wave function centered over the hub 
sites is always zero. Hence, electrons in the flat band remain localized 
around the rim sites.

For $\alpha = 1$, Eq. (2) reduces to the pseudospin-1 Dirac-Weyl 
Hamiltonian
$
H_{0}^{\mu}({\bf k}) = \hbar v_f \textbf{S} \cdot \textbf{k},
$
where $\textbf{S}=(S_x,S_y,S_z)$ are the standard spin-1 matrices.

{\bf Berry phase}:
The topological Berry phase for $\alpha = 0$ is simply $\pi$ which is 
independent of the valleys. 
For $\alpha > 0$, the $\alpha$-dependent Berry phase \cite{Illes} 
$\phi_B^{\lambda, \mu} $ in the conduction and valence bands is given by
\be \label{BP-D}
\phi_B^{\pm 1, \mu} =  \pi \mu \cos(2\phi) = 
\pi \mu  \Big(\frac{1-\alpha^2}{1+\alpha^2}\Big),
\ee
and for the flat band is given by
\be \label{BP-F}
\phi_B^{0,\mu} = - 2\pi \mu \cos(2\phi) = - 
 2\pi \mu \Big(\frac{1-\alpha^2}{1+\alpha^2}\Big).
\ee
Note that the Berry phase is different in the ${\bf K}$ and 
${\bf K}^{\prime}$ valleys except for $\alpha=1$.
The Berry phase is smoothly decreasing with increase of $\alpha$ and
becomes zero at $\alpha=1$.
Later, we will show how the Berry phase appears in quasienergy
gaps.

\section{Floquet eigensystem for $\alpha$-$T_3$ lattice}
We consider a circularly polarized electromagnetic radiation propagating 
perpendicular to the $\alpha$-$T_3$ lattice placed in the $x$-$y$ plane. 
The corresponding vector potential is given by
$ 
\textbf{A}(t) = A_0 (\cos \omega t \; \hat{\bf x} 
+ \nu \sin \omega t \; \hat{\bf y}),
$
where $ A_0 = E_0/\omega $ with $E_0$ is the amplitude of the electric field 
and $\omega$ is the frequency of the radiation. Also, $\nu = \pm 1$ denotes 
counter-clockwise and clockwise rotations 
of the circularly polarized light, respectively. 
The frequency of the driving is small compared to the bandwidth of the system.
The vector potential satisfies the
time periodicity: ${\bf A}(t+T) = {\bf A}(t) $ with the time-period $T=2\pi/\omega$.
The minimal coupling between the charge carrier and the electric field is obtained
through the Peierls substitution: 
$\hbar {\bf k} \rightarrow (\hbar {\bf k} - q {\bf A}(t))$ with $q=-e$ being the
electronic charge. 
The Hamiltonian for the coupling between the charge carriers and the electromagnetic
field can be written as
\be
H_1^{\mu\nu}(t) = \hbar \omega \beta [S_{-}^{\mu \nu} \e^{i\omega t }
+ S_{+}^{\mu \nu} \e^{-i\omega t }],
\ee
where the $3 \times 3 $ matrices are
$ S_{\pm }^{\mu \nu} = \frac{1}{2} [\mu S_x(\alpha) \pm i \nu S_y(\alpha)]$ and
the dimensionless parameter $\beta = e E_0 l_\omega/(\hbar \omega)$
characterizes the strength 
of the coupling between electromagnetic radiation and charge carrier with 
$l_\omega/a = 3 \pi t/\hbar \omega \gg 1 $ 
in the THz frequency regime. The dimensionless 
parameter $\beta $ is less than 1 for the typical intensity of lasers
available in the THz frequency regime.
In the semiclassical picture, 
$e E_0 l_\omega $ is the energy gained by the charge carrier while 
travelling a distance $l_\omega$ with the speed $v_f$ during one period of the
radiation. On the other hand, the charge carrier is dressed with the minimal 
photon energy $\hbar \omega $.

The total Hamiltonian of a charge carrier near the Dirac point in presence of
the electromagnetic radiation is 
$ H^{\mu \nu}({\bf k},t) = H_0^{\mu}({\bf k}) + H_1^{\mu \nu}(t)$ which is periodic
in time.
By Floquet theory, the solution of 
time-dependent Schrodinger equation 
\begin{equation} \label{Flo1}
i \hbar \frac{\partial}{\partial {t}}|\psi_{\eta}^{\mu \nu}(\textbf{k},t) \rangle = 
H^{\mu \nu} (\textbf{k},t)| \psi_{\eta}^{\mu \nu}(\textbf{k},t) \rangle
\end{equation}
is given by
\begin{equation} \label{Flo2}
|\psi_{\eta}^{\mu \nu}(\textbf{k},t) \rangle = 
\e^{-i\varepsilon_{\eta}^{\mu \nu}(\textbf{k}) t/\hbar}| 
\phi_{\eta}^{\mu \nu}(\textbf{k},t) \rangle.
\end{equation}
Here $|\phi_{\eta}^{\mu \nu}(\textbf{k},t) \rangle$ are the time-periodic 
$1\times 3 $ pseudo-spinors 
and $\varepsilon_\eta^{\mu \nu}(\textbf{k})$ are the corresponding quasienergies.
There are three independent quasienergy branches along with the three corresponding 
eigenstates indexed by $\eta = 1, 0, -1$.
Substituting Eq. (\ref{Flo2}) into Eq. (\ref{Flo1}), the time-periodic
spinor $|\phi_\eta^{\mu\nu}(\textbf{k},t)\rangle $ becomes the eigenstate of the
Floquet Hamiltonian $H_F^{\mu \nu} = H^{\mu \nu}({\bf k},t) - 
i \hbar \frac{\partial}{\partial t}$
with the eigenvalue $ \epsilon_\eta^{\mu \nu}(\bf{k})$:
\begin{equation} \label{Flo3}
\Big[ H^{\mu \nu}(\textbf{k}, t) - i \hbar\frac{\partial}{\partial t} \Big]
|\phi_{\eta}^{\mu \nu}(\textbf{k},t) \rangle 
= \epsilon_{\eta}^{\mu \nu} (\bf{k})| \phi_{\eta}^{\mu \nu} (\bf{k},t) \rangle.
\end{equation}

Multiplying a phase $\e^{-i m \omega t}$ with $m$ being an integer to Eq. (\ref{Flo2})
and substituting it back to Eq. (\ref{Flo3}), we obtain
\begin{equation} \label{Flo3a}
(H^{\mu \nu}(\textbf{k},t) - i \hbar\frac{\partial}{\partial t}) 
|\phi_{\eta}^{\mu \nu}(\textbf{k},t) \rangle =
(\varepsilon_{\eta}^{\mu \nu}(\textbf{k}) + m \hbar\omega)| 
\phi_{\eta}^{\mu \nu}(\textbf{k},t) \rangle.
\end{equation}
This is also an eigenvalue equation as Eq. (\ref{Flo3})  but with a shifted 
quasienergy 
$ \varepsilon_{\eta m}^{\mu \nu} (\textbf{k}) = 
\varepsilon_{\eta}^{\mu \nu}(\textbf{k}) + m \hbar \omega $.
Equations (\ref{Flo3}) and (\ref{Flo3a}) yield the same Floquet mode, 
with quasienergies differing by an integer multiple of photon energy $\hbar \omega$. 
Hence, the index $\eta$ corresponds to a whole class of solutions 
indexed by $\eta^{\prime} = (\eta,m), m=0, \pm1, \pm2,...$
having a discrete spectrum of quasienergies $\varepsilon_{\eta,m}(\textbf{k})$.
Thus, a given Floquet state has multiple quasienergy values repeating in
the intervals of $\hbar\omega$. For $\alpha$-$T_3$ lattice, we have three
independent values of quasienergy for a given momentum, which can be attributed
to the three independent eigenvalue equations $(\eta=1,0,-1)$.
Due to the infinite spectrum without physical distinguishability,
the quasienergies can also be confined to a reduced Brillouin zone in energy
space with $|\varepsilon_{\eta}^{\mu \nu}(\textbf{k})|<\hbar\omega/2$.

In order to calculate the quasienergies and the corresponding states of the 
Floquet Hamiltonian, we
consider the Fourier expansion of 
$|\phi_{\eta}^{\mu \nu} (\textbf{k},t)\rangle$ 
\be \label{Flo4}
|\phi_{\eta}^{\mu \nu}({\bf k},t) \rangle = 
\sum_{n = - \infty}^{\infty} \e^{-i n\omega t} |\chi_{\eta}^{n \mu \nu} ({\bf k})\rangle, 
\ee
which follows from the temporal periodicity of the Floquet mode.
Using Eq. (\ref{Flo4}), the time-dependent differential Eq. (\ref{Flo3})
reduces to the time-independent eigensystem problem as
\be \label{Flo-H}
\sum_m \Big[ H_{0F,mn}^{\mu \nu} +  H_{1F,mn}^{\mu \nu}  - 
\epsilon_{\eta}^{\mu \nu}({\bf k}) \Big] |\chi_{\eta}^{m \mu \nu} ({\bf k})\rangle
=0,  
\ee 
where the diagonal Floquet Hamiltonian in the Floquet basis is
\be
H_{0F,mn}^{\mu \nu} = [\hbar v_f {\bf S}(\alpha) \cdot {\bf k} + m \hbar \omega ]\delta_{mn}.
\ee  
and the off-diagonal interaction Hamiltonian
\be
H_{1F,mn}^{\mu \nu} = \hbar \omega \beta[ S_{-}^{\mu \nu} \delta_{m,n-1}
+ S_{+}^{\mu \nu} \delta_{m,n+1}]
\ee
couples various Fourier modes.
Thus, by Floquet matrix theory, we numerically compute the Floquet
quasienergies $ \varepsilon_{\eta}^{\mu \nu}$ in units of $\hbar\omega$
and the corresponding Floquet states 
$|\psi_{\eta}^{\mu \nu}(k,t) \rangle$
of the Floquet Hamiltonian $H_F^{\mu \nu}$.
The following parameters have been used in the numerical calculation:
$\omega = 2\pi \times 5 $ THz, $E_0 = 2 $ kV/cm, 
$v_f = 10^6 $ m/s and $\beta = 0.3$. 
Also, $\mu = \nu = + 1 $ are considered for all the plots unless 
otherwise stated. 

\subsection{Exact analytical expressions of quasienergies and band gap 
at the Dirac points}
First, we present exact analytical results of quasienergies and band gaps 
at the Dirac points. 
At the Dirac points (${\bf k} = 0$), the time-dependent Hamiltonian 
(in units of $\hbar \omega$) can be written as
\be
\bar H_1^{\mu\nu}(\tilde t) = \beta [S_{-}^{\mu \nu} e^{i\tilde t }
+ S_{+}^{\mu \nu} e^{-i\tilde t }],
\ee
where $\tilde t = \omega t$.
\begin{figure}[htbp]
\includegraphics[trim={0cm 0cm 0cm  0cm},clip,width=9cm]{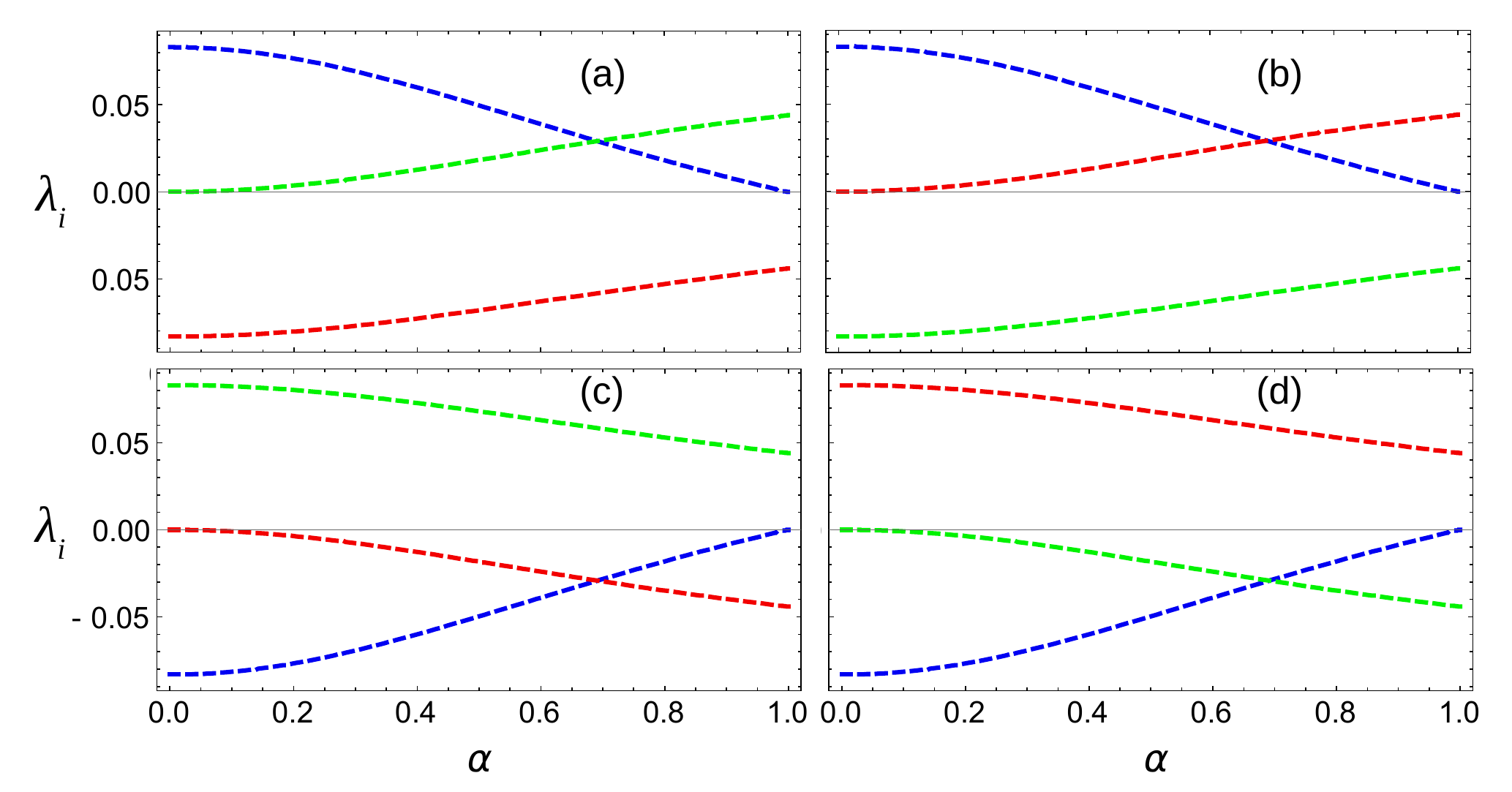}
\caption{Plots of quasienergies at the Dirac point vs $\alpha$ for various
combinations of $\mu$ and $\nu$:
(a) $\mu = \nu = 1$,
(b) $ \mu = \nu = - 1$,
(c) $\mu = 1, \nu = - 1$,
(d) $\mu = -1, \nu = 1$.
The green, red and blue dotted curves correspond to
$\lambda_1, \lambda_2$ and $\lambda_3$.}
\label{Dirac-quasiE}
\end{figure}
The corresponding Floquet Hamiltonian can be written explicitly as
\begin{equation}
\begin{aligned}
\tilde{H}_F^{\mu \nu }(\tilde{t}) = \mu \beta
\left[\begin{array}{ccc}-i\frac{\partial}{\partial \tilde{t}} & \cos \phi
\e^{-i\mu \nu \tilde{t}} & 0 \\\cos \phi \e^{i\mu \nu \tilde{t}} &
-i\frac{\partial}{\partial \tilde{t}} & \sin \phi \e^{-i\mu \nu \tilde{t}} \\
0 & \sin \phi \e^{i\mu \nu \tilde{t}} & -i\frac{\partial}{\partial \tilde{t}}
\end{array}\right].
\end{aligned}
\end{equation}
Let us define an unitary operator $\hat Q$ given by
\begin{equation}
\hat Q = e^{-i \mu \nu ({\it I} + S_z) \tilde{t}},
\end{equation}
where $ {\it I} $ is the $3 \times 3 $ identity matrix.
By performing the unitary transformation
$\hat Q^\dagger \tilde{H}_F^{\mu \nu}(\tilde{t}) \hat Q$, an effective
time-independent Floquet Hamiltonian is obtained
\begin{equation}
\tilde{H}_F^{\mu \nu } = \mu \left[\begin{array}{ccc}
- 2 \nu & \beta \cos \phi & 0 \\ 
\beta \cos \phi & - \nu &  \beta \sin \phi \\
0 & \beta \sin \phi & 0
\end{array}\right].
\end{equation}
The zero-momentum quasienergy spectra are given by
\begin{eqnarray}
\lambda_1 & = & - \mu\nu + \frac{2\mu}{\sqrt{3}}\sqrt{1 + \beta^2}\cos\Lambda  \label{exact1} \\
\lambda_2 & = & - \mu\nu - \frac{\mu}{3}\sqrt{1 + \beta^2}(\sqrt{3}\cos\Lambda + 3\sin\Lambda) \label{exact2}\\
\lambda_3 & = & - \mu\nu - \frac{\mu}{3}\sqrt{1 + \beta^2}(\sqrt{3}\cos\Lambda - 3\sin\Lambda) \label{exact3},
\end{eqnarray}
where $\Lambda = (1/3) $Arg$[- \xi + \sqrt{\xi^2 - 108(1+\beta^2)^3}]$ and
$\xi = 27 \nu\beta^2 \big(\frac{1-\alpha^2}{1+\alpha^2}\big)$ with 
Arg$[z]$ gives the argument of the complex number $z$. The corresponding normalized Floquet states are given by
\begin{eqnarray}
| \psi^{\mu\nu}(\tilde{t}) \rangle = \frac{\e^{-i(\lambda_i +\mu \nu) \tilde{t}}}{\sqrt{1+\beta^2 
\big(\frac{\cos^2\phi}{(2\mu\nu+\lambda_i)^2}+\frac{\sin^2\phi}{\lambda_i^2}\big)}} 
\left(\begin{array}{c}
\frac{\beta \cos\phi}{2\mu\nu+\lambda_i} \e^{-i\mu\nu\tilde{t}} 
\\ 1
\\ \frac{\beta \sin\phi}{\lambda_i} \e^{i \mu \nu \tilde{t} }
\end{array}\right). \nn
\end{eqnarray} 
The parameter $\xi$ can be expressed in terms of the Berry phase given by 
Eqs. (\ref{BP-D}) and (\ref{BP-F}). Thus, the quasienergy is directly
related to the Berry phase acquired during a cyclic motion of the charge
carriers in presence of a circularly polarized radiation.

\begin{figure}[htbp]
\includegraphics[trim={0cm 0cm 0cm  0cm},clip,height=2.85cm,width=8cm]{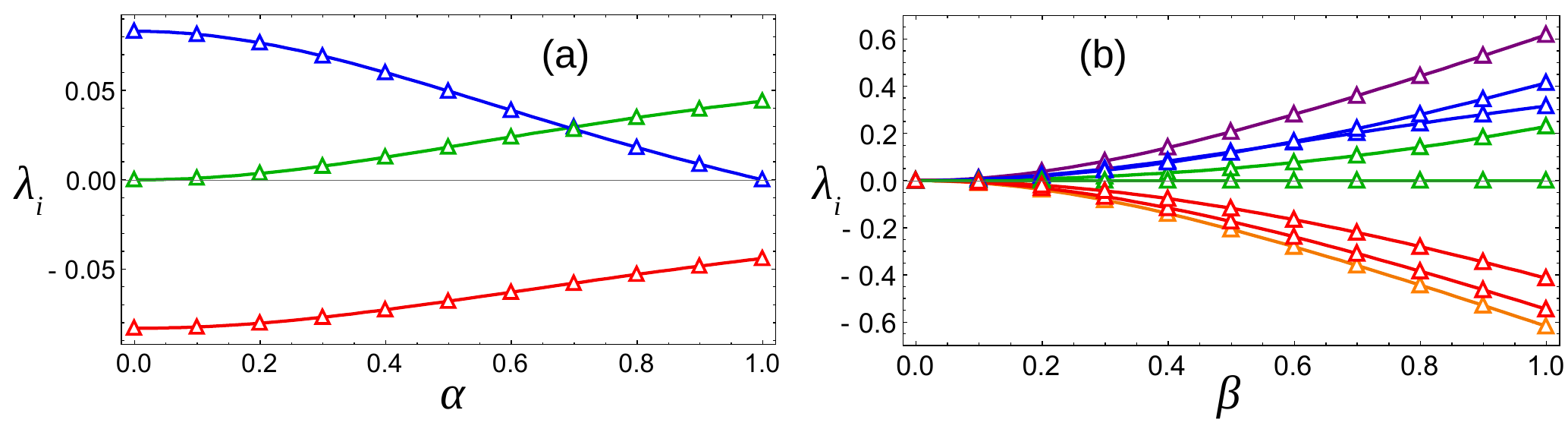}
\caption{Plots of variation of exact quasienergy at the Dirac point with
(a) $\alpha$ for $\beta=0.3$ and (b) $\beta$ for different  values of $\alpha$.
Exact numerical results are denoted by the triangles.
In Fig. (b), the solid green, blue and red lines
are for $\alpha=1$, the dotted lines for $\alpha=0.5$ and the purple-orange pair is
for $\alpha=0$.}
\label{Figexact}
\end{figure}

The three Floquet quasienergy branches may be labelled as $(i,m)$, 
where $i=1,2,3$ represent three branches and $m$ the Floquet index.
The corresponding quasienergy $\lambda_{i,m} = \lambda_i + m$. 
The quasienergies of the three branches in the first energy Brillouin zone 
are given by $\lambda_{1} + \mu (\nu-1), \lambda_{2} + \mu(\nu+1)$ 
and $\lambda_3 + \mu\nu$. 
The three-fold degeneracy at the Dirac point is simply the limiting case 
($\beta \rightarrow 0$) of these quasienergies.  The variation of these quasienergies
with $\alpha$ for $ \beta = 0.3$ is shown in Fig. \ref{Dirac-quasiE}.
Figure \ref{Dirac-quasiE} displays the photoinduced valley and electron-hole symmetry
breaking at the Dirac point for $0<\alpha<1$.
The quasienergy variations for the pair having same value of $\mu \nu$ (i.e. $\mu\nu=1$
[Fig. \ref{Dirac-quasiE}(a) and \ref{Dirac-quasiE}(b)] or 
$\mu \nu = -1$ [Fig. \ref{Dirac-quasiE}(c) and \ref{Dirac-quasiE}(d)]) are 
identical to each other apart
from the interchange of branches. Also, the quasienergy structure for the cases of
$\mu\nu = 1$ and $\mu\nu=-1$ are inverted copies of each other for $0<\alpha<1$.
This implies that spectrum undergoes a flipping on- (i) switching between valleys
$ \textbf{K}$ and $\textbf{K}^{\prime}$ for a given sense of circular polarization and (ii)
changing sense of rotation of the polarization for a given valley. The flipping of
quasienergies is trivially symmetric for graphene ($\alpha$=0) and dice lattice ($\alpha$=1)
on switching of valleys or polarization. 


For $\alpha=0$, the quasienergies within the first energy BZ 
are obtained as
\begin{equation} \label{oka}
\lambda_\pm = \pm \frac{1}{2}(\sqrt{4\beta^2+1} - 1).
\end{equation}
The same results are obtained by Oka and Aoki \cite{Oka} for irradiated graphene.
On the other hand, the quasienergies for the dice lattice ($\alpha=1$) obtained from 
Eqs. (\ref{exact1},\ref{exact2},\ref{exact3})  are 
$\lambda_0 = 0$ and 
\begin{equation} \label{dey}
\lambda_\pm = \pm(\sqrt{\beta^2 + 1} - 1).
\end{equation}
Equations (\ref{oka}) and (\ref{dey}) can be combined to write a general
form for quasienergy at the Dirac point as
\begin{eqnarray} \label{gap}
\lambda_\pm (S) = \pm(\sqrt{\beta^2 + S^2}- S),
\end{eqnarray}
where $S$ is the pseudospin of the underlying lattice. 
The energy gap at the Dirac point for the pseudospin $S$ is 
$ \Delta_S = \lambda_+(S) - \lambda_-(S)  = 2(\sqrt{\beta^2 + S^2}- S)$.  
The energy gap for graphene is 
$\Delta_{S=\frac{1}{2}} = (\sqrt{4 \beta^2 + 1} - 1)$ and that for dice lattice
is $\Delta_{S=1} = 2(\sqrt{\beta^2 + 1} - 1)$.
It can be easily checked from Fig. 3(a) as well as from
Eq. (\ref{gap}) that $ \Delta_{S=1} < \Delta_{S=\frac{1}{2} }$.
The quasienergy gap at the Dirac point for graphene is higher than that
of dice lattice.
Thus, the flat band has a shielding effect on the dipole coupling
between the electron-photon levels.

In Fig. \ref{Figexact}, we show the variation of the three quasienergy branches 
with $\alpha $ and $\beta$. The colour labeling of Fig. \ref{Figexact} is the same as that of 
Fig. \ref{Dirac-quasiE}(a).
Figure \ref{Figexact}(a) shows that our numerical results
match very well with the exact results. The quasienergy at the Dirac point increases
with the field strength $\beta$ seen at Fig. \ref{Figexact}(b).

\subsection{Floquet quasienergy branches and band gaps for large values of
momentum}
\begin{widetext}
\begin{figure*}[htbp]
\includegraphics[trim={3cm 0 3cm  11cm},clip,width=17cm]{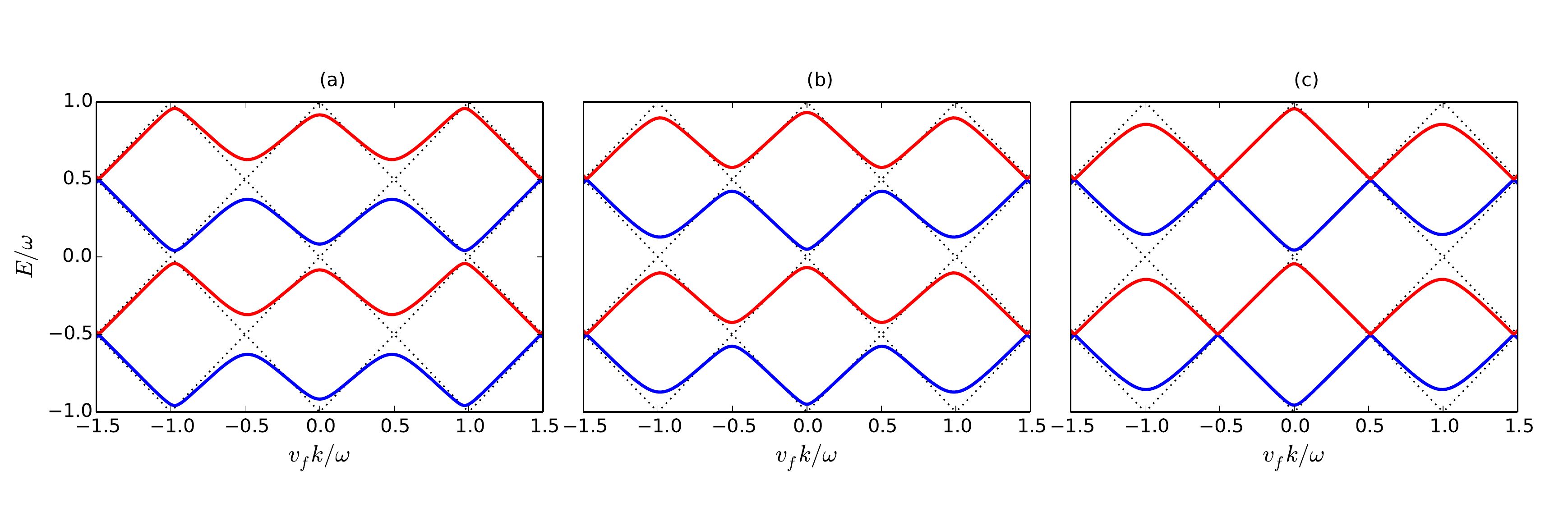}
\caption{Floquet quasienergy bands for different values of $\alpha$:
(a) $\alpha = 0$, (b) $\alpha = 0.5$, (c) $\alpha = 1$.} 
\label{Figexact1}
\end{figure*}
\end{widetext}

\begin{figure}[htbp]
\includegraphics[trim={1.5cm 0 2cm  0},clip,width=8cm]{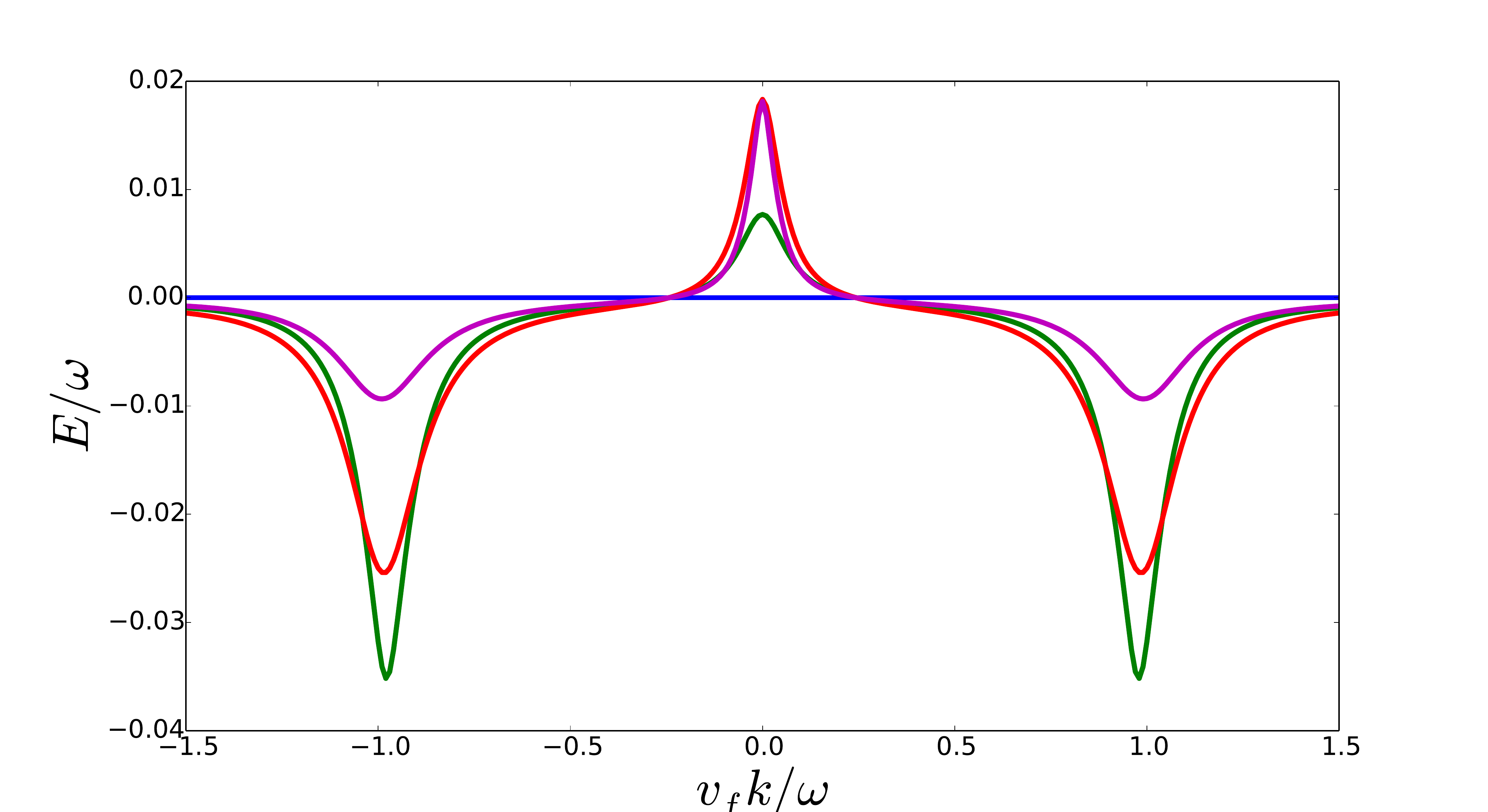}
\caption{Quasienergy band for $\eta=0$ at different values of
$\alpha$: (i) $\alpha=0.3$ (green), (ii) $\alpha=0.5$ (red),  (iii) $\alpha=0.8$ (purple)
and $\alpha=1$ (blue).}
\label{exact-flat}
\end{figure}

In this subsection, we present the results obtained by solving the low-energy 
Floquet Hamiltonian numerically and display the quasienergy band structure 
within the first two energy Brillouin zones in Fig. \ref{Figexact1} for 
$\alpha = 0, 0.5$ and 1. The dotted lines indicate the spectrum for zero 
intensity of radiation, which are identical for all values of $\alpha$.
First of all, the quasienergies of the  $\alpha$-$T_3$ lattice pertaining 
to different $\eta$ satisfy 
$\varepsilon_{1}^{\mu \nu}({\bf k}) = - \varepsilon_{-1}^{\mu \nu}(\textbf{k})$
for $\alpha = 0, 1$ and 
$ \varepsilon_{\eta}^{\mu \nu}(\textbf{k}) = \varepsilon_{\eta}^{\mu \nu}(- \textbf{k})$
for $0 \leq \alpha \leq 1$.
The known results of graphene ($\alpha=0$) are reproduced in Fig. \ref{Figexact1}a.
For better visualization, the quasienergy band for $\eta=0$ is shown separately 
in Fig. \ref{exact-flat}. 
The quasienergy branch corresponding to the flat band becomes dispersive mainly 
around the Dirac and even-photon resonant points for $ 0 < \alpha < 1$.
This is due to the fact that the flat band states are
dressed with integral number of photons in the vicinity of these resonant points, 
which allows them to mix with dressed conduction and valence band states. The
mixing results in dispersion due to shifting of energies of the erstwhile non-dispersive states.
On the other hand, the band does not undergo any significant
modification at odd-photon resonances, as it cannot be dressed with half-integral number of photons. 
The dispersion gets completely wiped out at $\alpha=1$.
The height of the spikes of the dispersion decreases with increases of the momentum. 
The band structure gets inverted about the $ k$ axis on 
changing the rotation of electric field vector of the circularly polarized light. 
The band remains flat for all values of $\alpha$ when applied radiation 
is linearly polarized. 
It is to be noted that there is no splitting in the flat band since 
it does not have any partner band.

\begin{figure}[htbp]
\includegraphics[trim={0cm 0 0cm  0},clip,width=7cm]{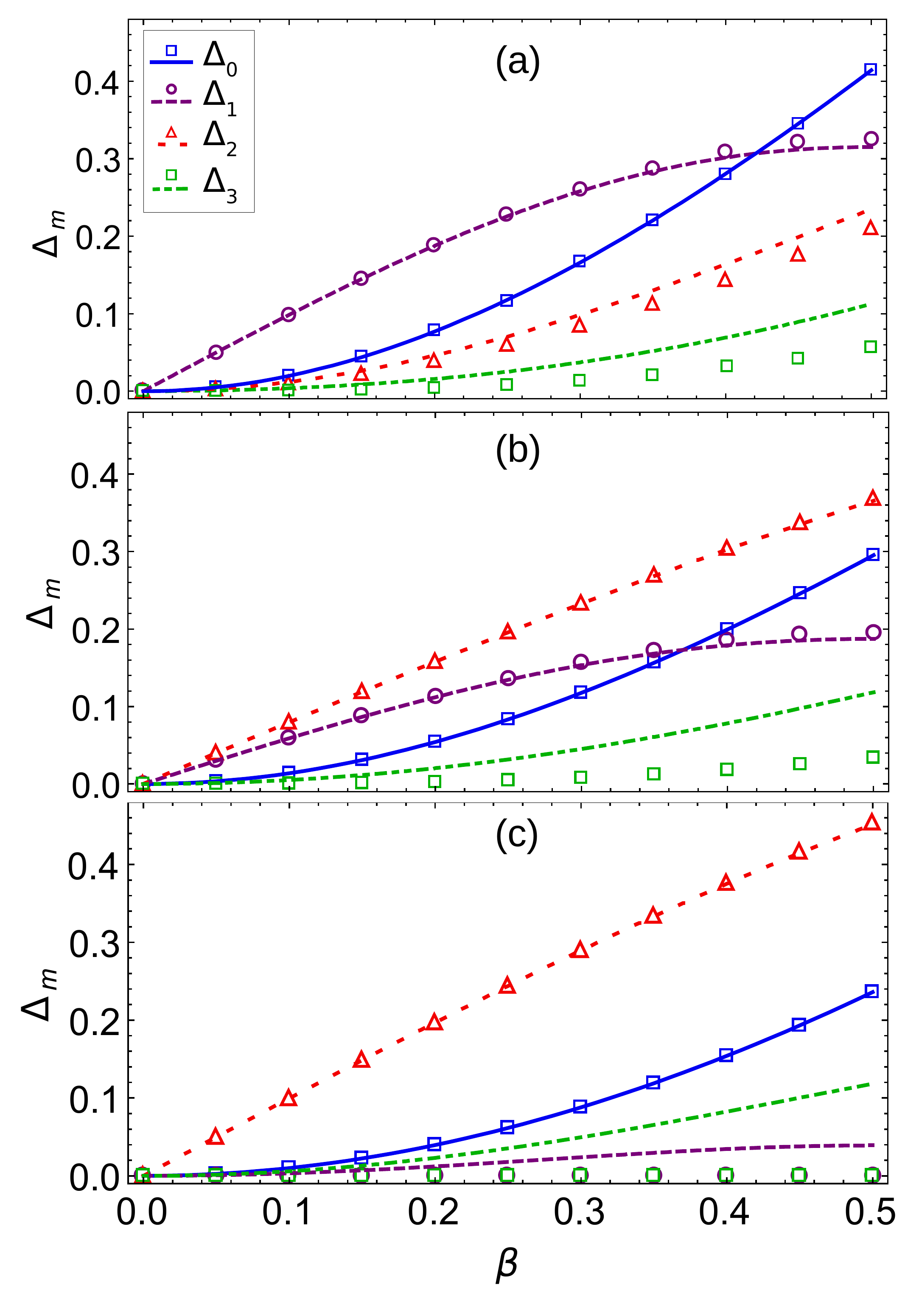}
\caption{Plots of variation of gaps $\Delta_m$ with $\beta$ for
(a) $\alpha=0$, (b) $\alpha=0.5$ and (c) $\alpha=1$.}
\label{gap-beta}
\end{figure}
The gaps between the bands ($\eta = \pm 1) $) open up at $k=0 $ and at
$k_m = m \omega/2 v_f$ with $m = \pm 1, \pm 2,...$.
The gap at $k_m$ arises due to the AC Stark splitting occurrs due to 
the multiphoton resonances \cite{Gupta,Home,Grifoni,Faisal,Hsu,Acosta}. 
There is a set of Bloch states lying on a circle in the vicinity of the Dirac point $k$-space 
with radius $k_m$ such that energy difference between the bands is 
$m$ multiples of photon energy: $2v_f k_m = m \omega$.
On illumination, new electron-photon states with energy 
$E_\lambda = \hbar v_f k_m + N_\lambda \hbar \omega$ 
($\lambda$ band with $N_\lambda $ photons) and 
$E_{\lambda^{\prime} } = - \hbar v_f k_m + N_{\lambda^{\prime} } \hbar \omega$
with $\lambda^{\prime} \neq \lambda $ 
($\lambda^{\prime}$ state with $N_{\lambda^{\prime}} $ photons) are formed.
When $E_\lambda = E_{\lambda^{\prime} }$ i.e $N_\lambda - N_{\lambda^{\prime} } = m \hbar\omega$, 
the degenerate levels split due to the coupling between the electron and the radiation 
field and the gap opens up at $k_m$. All the gaps tend to diminish at 
higher values of momentum.

\begin{figure}[htbp]
\includegraphics[trim={0cm 0 0cm  0},clip,width=7cm]{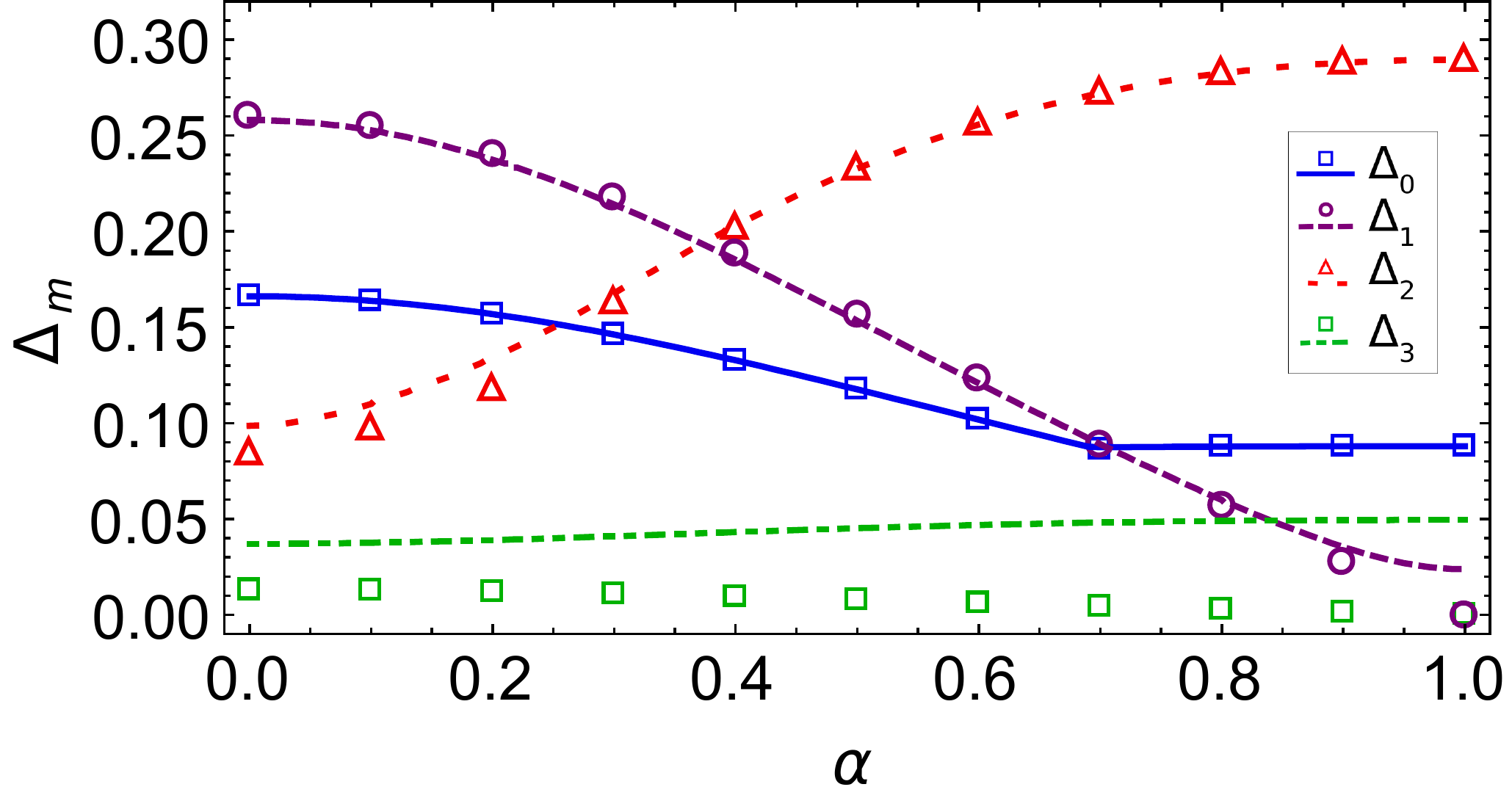}
\caption{Plots of variation of gaps $\Delta_m$ with $\alpha$ for $\beta=0.3$.}
\label{gap-alpha}
\end{figure}

Using the rotating wave approximation (see Appendix),  
the approximate quasienergies for $\alpha=0$ (for any integer $m$)
and $\alpha=1$ (for even $m$) are, respectively, given by    
\begin{eqnarray}
(\lambda_{\pm})_{\alpha=0} & = & \pm \frac{\beta }{2} |J_{m+1}(2\beta) - J_{m-1}(2\beta)| 
\label{a0} \\
(\lambda_{\pm})_{\alpha=1} & = & \pm \frac{\beta }{2} |J_{m/2+1}(\beta) - J_{m/2-1}(\beta)|
\label{a1}.
\end{eqnarray}
From the above expressions, we can see that $(\lambda_{\pm})_{\alpha=0}$ and
$(\lambda_{\pm})_{\alpha=1}$ are proportional to the difference between two 
consecutive integral and even ordered Bessel functions, respectively. The magnitude 
of the gaps is strongly affected by the argument of the Bessel functions which, 
for graphene, is twice as that of dice lattice. For $\beta \ll 1$, the asymptotic 
forms of quasienergy gaps are $ (\Delta_m)_{\alpha=0} \sim \beta^m $ and
$ (\Delta_m)_{\alpha=1} \sim (\beta/2)^{m/2} $. For weak fields, $(\Delta_1)_{\alpha=0}$ 
and $(\Delta_2)_{\alpha=1}$ vary linearly with $\beta$.
The variation of the gaps $\Delta_m$ with $\beta $ and $ \alpha $ are shown in
Fig. \ref{gap-beta} and Fig. \ref{gap-alpha}, respectively. 
The curves represent the numerical results while their corresponding markers
represent the results obtained from exact analytical
expressions ($\Delta_0$) and rotating wave approximation ($\Delta_1, \Delta_2, \Delta_3$).
All the gaps increase monotonically with $\beta$. 
Moreover, $\Delta_0$ (solid blue) and $\Delta_1$ (dashed purple) get reduced at higher $\alpha$. 
In contrast, $\Delta_2$ (dotted red) is found to increase with $\alpha$. 
The gap $\Delta_3$ (dashed-dotted green) increases very slowly with $\alpha$.
The splitting at even-photon resonant points is affected by the intervention of the flat band
dressed with integral number of photons. The interplay
of the three bands results in a increase in magnitude of the gap as $\alpha$ increases.
We see that the agreement between numerical and analytical results does not hold
good at higher momentum.

\section{Time-averaged energies and density of states}
In this section, we discuss about time-averaged quantities such as 
mean energy and the corresponding density of states. 
The mean energy is a single-valued quantity, which is independent 
of the choice of the quasienergy of the Floquet state.
The mean energy helps to understand whether the Floquet
states are occupied or unoccupied \cite{Gupta,Faisal,Hsu}. 
For example, the Floquet states having lower mean quasienergy 
will be accommodated first. Also, the mean quasienergy can be used 
to characterize whether the state is electron-like or hole-like \cite{Wu}. 

{\bf Time-averaged  energies}:
The expectation value of the Hamiltonian in a Floquet state is a 
periodic function of time. This helps us to formulate the energy averaged 
over a full cycle of a periodic driving and is given by
\begin{equation}
\bar{E_{\eta}^{\mu\nu}}(\textbf{k}) =
\frac{1}{T}\int_{0}^{T}\langle\psi_{\eta}^{\mu \nu}(\textbf{k},t)| H^{\mu \nu}(\textbf{k},t)|
\psi_{\eta}^{\mu \nu}(\textbf{k},t) \rangle dt.
\end{equation}
Incorporating the Fourier series (Eq. (\ref{Flo4})) into the above equation, we get
\begin{equation}
\bar{E}_{\eta}^{\mu\nu}(\textbf{k}) = \varepsilon_{\eta}^{\mu \nu}(\textbf{k}) +
\sum_{n = -\infty}^{\infty} n \hbar \omega \langle \chi_{\eta}^{n \mu \nu}(\textbf{k})|
\chi_\eta^{n\mu}(\textbf{k}) \rangle.
\end{equation}
Hence, the averaged energy can be viewed as the weighted average of energies
possessed by the Fourier harmonics of the Floquet modes.
\begin{widetext}
\begin{figure*}[htbp]
\includegraphics[trim={3cm 0cm 3cm  11cm},clip,width=17cm]{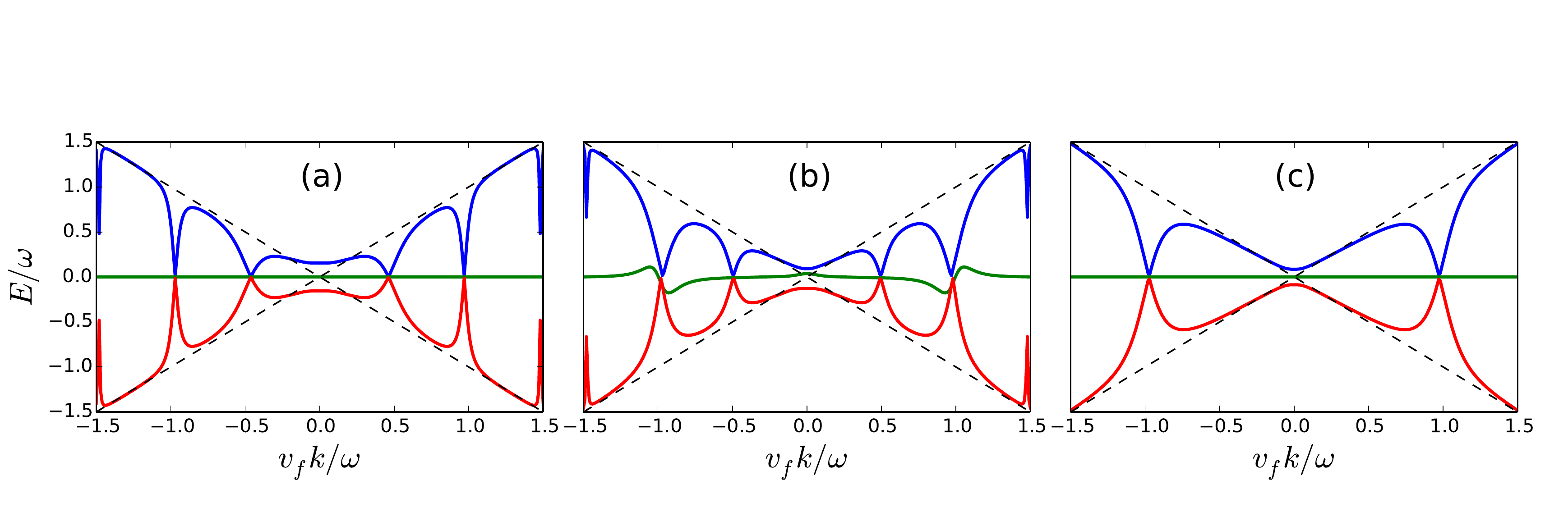}
\caption{Time-averaged energy of the three Floquet branches for 
(a) $\alpha = 0$, (b) $\alpha = 0.5$, (c) $\alpha = 1$.}
\label{mean-energy}
\end{figure*}
\end{widetext}
The time-averaged energy (around ${\bf K}$ valley) corresponding to the Floquet states of the
three branches for different values of $\alpha$ are shown in Fig. \ref{mean-energy}.
The blue, green and red bands represent the quasielectron, flat band and 
quasihole states respectively.
For $0 < \alpha < 1$, the mean energy band structure around ${\bf K}^{\prime}$ can be 
obtained by simply inverting the same for ${\bf K}$ valley. Due to absence of inversion 
symmetry of the band structure for $0 < \alpha < 1$, the valley degeneracy is broken.
As $\beta \rightarrow 0$, we obtain the field-free Dirac cones shown by the dotted lines.
For $0 \leq \alpha < 1$, the mean energy goes to zero near one-photon and two-photon resonant 
points due to crossover between quasielectron and quasihole states \cite{Wu}. 
The vanishing of mean energies at these  resonant 
points is also observed for $\alpha<1$. But, dice lattice has a non-zero mean 
energy near $one$-$photon$ resonance. This can be attributed to the fact that 
the gap at $one$-$photon$ resonance becomes vanishingly small and mimics 
the radiation-free case for $\alpha=1$.
A finite gap exists at the Dirac point for all values of $\alpha$.
The three-fold degeneracy at the Dirac point is lifted by the radiation.
Careful examination reveals that the symmetric nature of 
$\eta = \pm 1$ bands (electron-hole symmetry) is slightly disrupted near the 
Dirac point for $0 < \alpha <1$. The symmetry is restored at $\alpha=1$.
A distortion occurs in the mean energy spectrum of the flat band at
$k=0, \pm \omega/v_f$, similar to that obtained in the quasienergy spectrum.
The distortions flatten out at $\alpha=1$.

{\bf Time-averaged density of states}:
The time-averaged density of states over a driving cycle is defined as 
\begin{equation}
D(E) = g_s \sum_{n,\eta,\mu, \textbf{k}}^{}\langle\chi_\eta^n(\textbf{k})| 
\chi_\eta^n(\textbf{k})\rangle 
\delta\big(E-(\varepsilon_\eta(\textbf{k})+n \hbar \omega)\big)
\end{equation}	
Here, the factor $g_s=2$ appears due to the spin degeneracy. On converting 
the sum over $\textbf{k}$ to integral, i.e 
$\sum \limits_{\textbf{k}}^{}\rightarrow \frac{1}{(2\pi)^2} 
\int_{}^{\textbf{k}} d^2 \textbf{k}$ and using the azimuthal symmetry of 
quasienergy band structure for circularly polarized light, we get the 
density of states per unit area as
\begin{equation}
g(E) = D_0\sum\limits_{\eta, n, \mu}^{}\int_{0}^{\infty} 
\langle\chi_\eta^{n\mu}(\tilde{k})|\chi_\eta^{n\mu}(\tilde{k})\rangle 
\delta(\tilde{\varepsilon}-(\tilde{\varepsilon}_\eta^\mu(\tilde{k})+n)) 
\tilde{k}d\tilde{k}
\end{equation}
where $D_0 = g_s \omega/(2\pi \hbar v_f^2) = 
1.515 \times$ 10$^{13}$ meV$^{-1}$ m$^{-2}$. 
$\tilde{\varepsilon}$ and $\tilde{\varepsilon}_\eta^\mu(\tilde{k})$ 
are the dimensionless quasienergies such that 
$E = \tilde{\varepsilon} \hbar\omega$.
Using the property of the Dirac-delta function, the above integral 
can be further simplified as
\begin{equation}
g(E) = D_0 \sum_{\gamma, \tilde k_{i}^{(\gamma)}(\tilde{\varepsilon} )} 
\frac{\langle \chi_{\eta}^{n\mu}(\tilde k_{i}^{(\gamma)}(\tilde{\varepsilon})| 
\chi_\eta^{n\mu}(\tilde k_{i}^{(\gamma)}(\tilde{\varepsilon}) \rangle 
\tilde k_{i}^{(\gamma)}(\tilde{\varepsilon}) }
{|\tilde{\varepsilon}_{\eta}^{\prime}(\tilde k_{i}^{(\gamma)} 
\tilde{\varepsilon}) |}
\end{equation}
where $\gamma = \{n, \eta, \mu \} $ is a set of quantum numbers and
$\tilde k_{i}^{(\gamma)}(\tilde{\varepsilon} )$ is the 
$i$-th positive root of 
$\tilde{\varepsilon} - (\tilde{\varepsilon}_\eta^\mu(\tilde{k})+n) =0$ for a given $n$.

\begin{figure}[htbp]
\includegraphics[trim={3cm 0 8cm  0},clip,width=15cm]{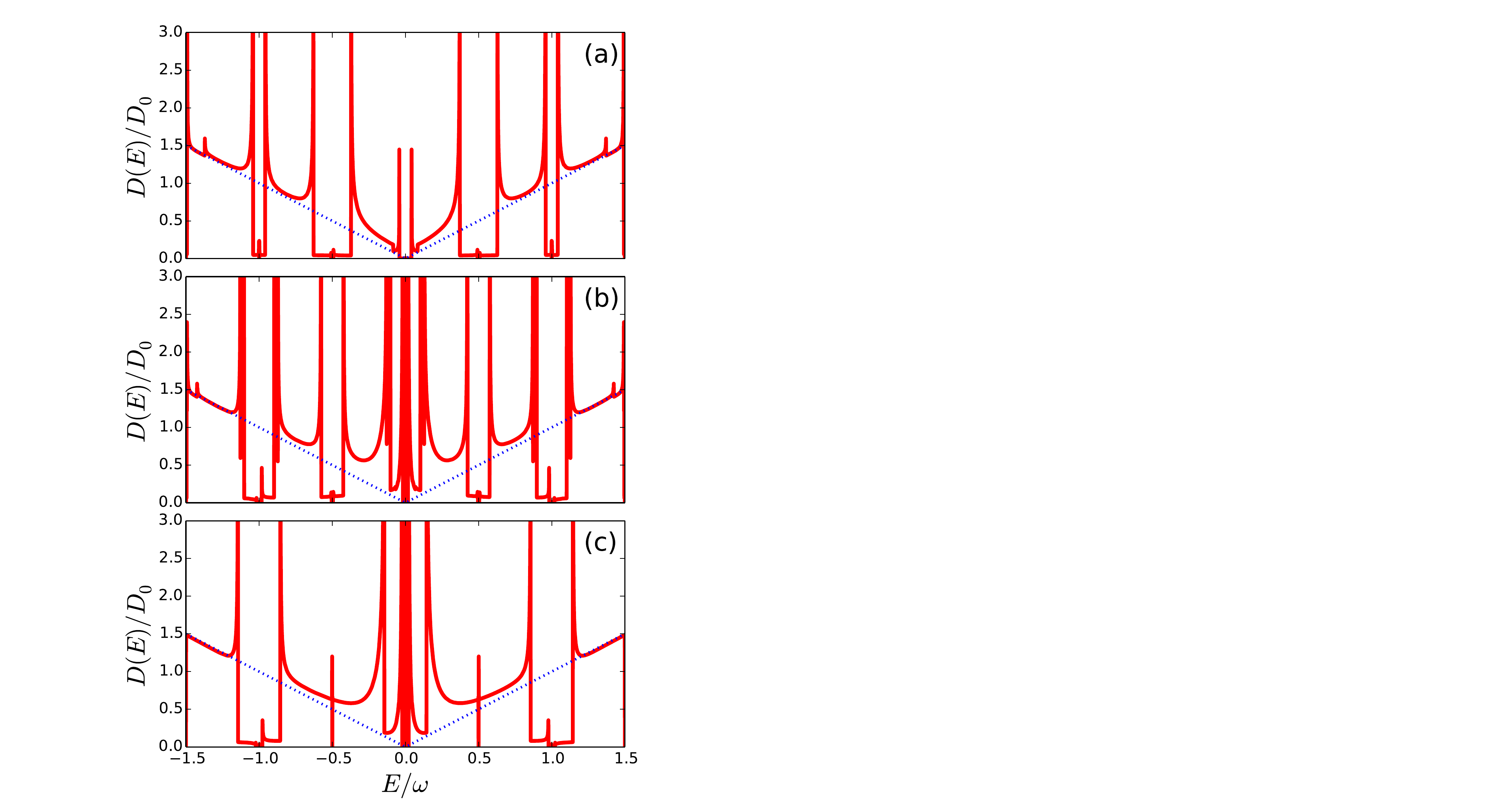}
\caption{Time-averaged density of states of quasielectron and quasihole  
states for (a) $\alpha = 0$, 
(b) $\alpha = 0.5$, (c) $\alpha = 1$.}
\label{mean-dos}
\end{figure}

\begin{figure}[htbp]
\includegraphics[trim={1cm 0 1.5cm  0},clip,width=8cm]{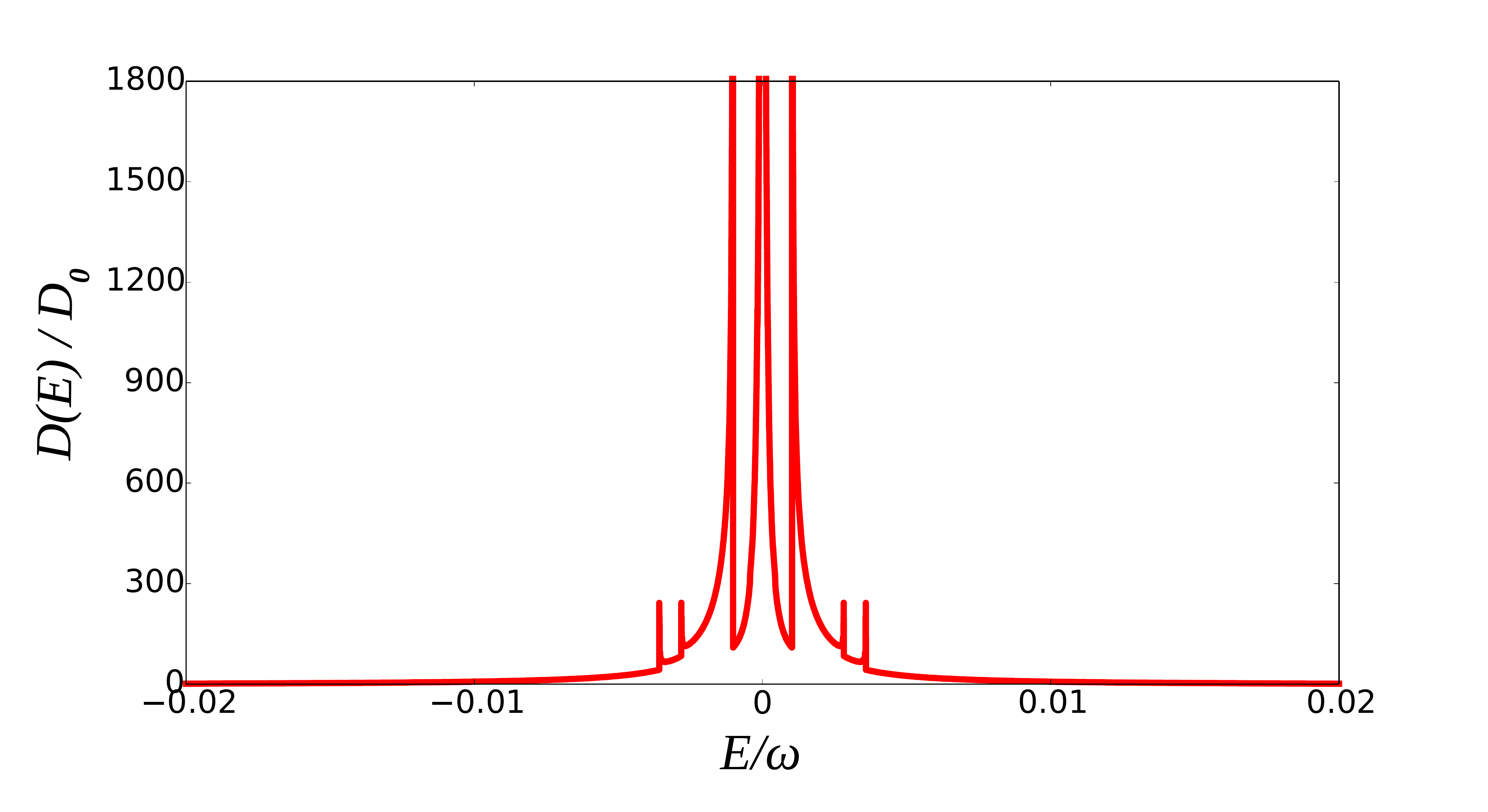}
\caption{Time-averaged density of states of quasiflat band around the Dirac point
for $\alpha=0.5$.}
\label{mean-dos-flat}
\end{figure}
Time-averaged density of states of electronlike and holelikes quasienergy bands
for three different values of $\alpha$ is shown in Fig. \ref{mean-dos}.
Figure \ref{mean-dos}(a) shows the DOS for $\alpha=0$, which is
similar to the results obtained by other groups \cite{Oka,Wu}.
The peaks represent the Van Hove singularities occuring due
to the extrema in the quasienergy band structure. Apart
from large peaks, there are spikes around $E = \pm \omega/2$  and
$E= \pm \omega$ with vanishingly small DOS. This is because of the photoinduced
gaps at the boundaries of the energy Brillouin
zones. A small but finite contribution of DOS in these
energy ranges  appears due to closing of gaps at higher momenta.
Additional peaks are born at the Dirac point for finite $ 0 < \alpha \leq 1 $ as
seen in Figs. \ref{mean-dos}(b) and \ref{mean-dos}(c). 
The separation between the peaks centred around $ E = \pm \omega/2$ decreases with
$\alpha$, while that around $E = \pm \omega$  increases with $\alpha$. 
This is related to the fact that $\Delta_1 (\Delta_2) $ decreases (increases) 
with $\alpha$. 

The time-averaged DOS for the flat band quasienergy around the Dirac point 
is shown in Fig. \ref{mean-dos-flat}.
Since the regions between two consecutive even-photon resonant points are 
predominantly flat, a large peak appears at zero energy. The dispersion at even-photon
resonant points in the flat band lead to occurence of additional peaks symmetrically 
placed around around the peak at zero energy. Similar feature in the DOS is repeated 
at energies equal to integral multiples of photon energy.
In dice lattice, only central peaks are present at $N\hbar \omega$ ($N$ being the integer)
due to absence of dispersion in the flat band.

\section{Summary and Conclusion}
We have investigated the Floquet quasienergy spectrum numerically and analytically 
for the $\alpha$-$T_3$ lattice driven by circularly polarized radiation. 
Exact analytical expressions of the quasienergy at
the Dirac points for all values of $\alpha$ and field strength are provided.
The band gap at the Dirac point appears due to the circularly
polarized radiation for all values of $\alpha$. 
The quasienergy gap at the Dirac point decreases with the increase of $\alpha$.  
Within the rotating wave approximation, we are able to get
approximate expressions of quasienergy at single-photon and multi-photon resonant
points. Approximate results match very well with the numerical results based on Floquet method.
The expressions reveal that the quasienergy is directly related to the
Berry phase acquired during a cyclic motion driven by the rotating electric field.
The valley symmetry is broken due to different Berry phase for different valleys  
for $0 < \alpha <1$. 
The quasienergy flat band remains dispersionless in presence of radiation for dice
lattice.  
However, dispersive spikes appear in and around the Dirac and even-photon resonant
points for $0<\alpha<1$.
The mean energy is non-vanishing around single-photon resonance point for dice lattice 
unlike $\alpha<1$. In contrast to graphene, we find that additional peaks appear 
in the time-averaged density of states at the Dirac point for $0<\alpha \leq 1$.
The pattern of the DOS near the single-photon and two-photon resonant points
varies significantly with $\alpha$.

Floquet-Bloch states on the surface of a topological insulator have been
observed using time- and angle-resolved photoemission spectroscopy (TrARPES)\cite{TI-exp}.  
There is a possibility that the quasienergy band structure of the $\alpha$-T${}_3$ lattice 
may be probed using TrARPES on subjecting the lattice to intense microwave pulses 
perpendicular to the lattice plane.
The variation in quasienergy band gaps 
with $\alpha$ may be observed by modulating the phase of one of the three counter-propagating 
laser beams.
Similarly, the quasienergy band structure for $\alpha=1/\sqrt{3}$  may be verified
by devising suitable means to irradiate Hg${}_{1-x}$Cd${}_x$Te  quantum wells.
The radiation-dressed band structure of such systems may open up doors for new opto-electronic devices.

\section{Acknowledgement}
We would like to thank Firoz Islam and Sonu Verma for useful discussions.

\section{APPENDIX}

\subsection{Analytical results within rotating wave approximation}
In this appendix, we shall derive analytical expressions of the 
quasienergy branches within rotating wave approximation \cite{Wu}.  
The analytical expressions help us to understand the Berry phase
dependency of the quasienergy bands and band gaps.
 
The time-periodic Hamiltonian $H^{\mu \nu}({\bf k},t)$ can be transformed 
in the basis formed by the eigenvectors of the low-energy Hamiltonian 
$H_{0}^{\mu}(\textbf{k})$ with the help of the unitary operator 
$\hat U_{\textbf{k}}$ given by
\begin{equation}
\hat U_{\textbf{k}} = \frac{1}{\sqrt{2}} 
\left(\begin{array} {ccc}
\mu \cos\phi e^{-i \mu \theta_\textbf{k}}  & \sqrt{2} \sin\phi e^{- i \mu \theta_\textbf{k}} & 
\mu \cos\phi e^{-i \mu \theta_\textbf{k}}\\ 
1 & 0 & -1 \\ 
\mu \sin \phi e^{i \mu \theta_\textbf{k}} & -\sqrt{2}\cos\phi e^{i \mu \theta_\textbf{k}} & 
\sin\phi e^{i \mu \theta_\textbf{k}}
\end{array}\right) \nn.
\end{equation}
The transformed Hamiltonian 
$ \hat U_\textbf{k}^\dagger H^{\mu \nu} (\textbf{k},t) \hat U_\textbf{k} 
= \tilde  H^{\mu \nu} (\textbf{k},t)$ reads as
\begin{equation}
\tilde H^{\mu \nu} (\textbf{k},t) = 
\hbar \omega \Big[\frac{v_f k}{\omega}  S_z + 
\tilde{H}_{\rm intra}(t) + \tilde{H}_{\rm inter}(t) \Big],
\end{equation}
where $ \tilde H_{\rm intra}(t) =  \beta 
(\cos \mu \theta_k \cos \omega t + 
\mu \nu \sin \mu \theta_k \sin \omega t)S_z$ 
with $S_z$ being the $z$-component of the spin-1 matrix and
\begin{equation}
\begin{aligned}
& \tilde H_{\rm inter}(t) =  \frac{i f(t)}{\sqrt{2}} 
& \left[\begin{array}{ccc}
0 &  \sin 2\phi & \sqrt{2} \mu \cos 2\phi \\ 
- \sin 2 \phi & 0 &  \sin 2\phi \\
-\sqrt{2} \mu \cos 2\phi & -  \sin 2\phi & 0
\end{array}\right]
\end{aligned} 
\end{equation}
where $ f(t) = \beta (\nu \cos \mu \theta_k \sin \omega t - 
\mu \sin \mu \theta_k \cos \omega t )$.

The Schrodinger equation is then given by
\begin{equation} \label{A-Schro}
\Big[ \frac{i}{\omega} \partial_{t} - 
\big(\frac{v_f k}{\omega} S_z + \tilde{H}_{\rm intra}(t) + \tilde{H}_{\rm inter}(t)\big) \Big]
|\psi_{\eta}({\bf k}, t) \rangle =0.
\end{equation}

We solve the Schrodinger equation by omitting the interband term 
$\tilde{H}_{\rm inter}(t)$ and get the following solutions:
\begin{equation}
\begin{aligned}
& |\psi_{+1}^{(0)}({\bf k}, t)\rangle = 
e^{-i v_f k t} u (\theta_{\textbf{k}},t) 
\left(\begin{array}{ccc}
1 \\ 0 \\ 0
\end{array}\right) \\
& |\psi_{-1}^{(0)}({\bf k}, t)\rangle = 
e^{i v_f k t} u^* (\theta_{\textbf{k}},t) 
\left(\begin{array}{ccc}
0 \\ 0 \\ 1 
\end{array}\right) \\
& |\psi_{0}^{(0)}({\bf k}, t)\rangle = 
\left(\begin{array}{ccc}
0 \\ 1 \\ 0 
\end{array}\right)
\end{aligned}
\end{equation}
where 
$ u \big(\theta_{\textbf{k}},t) = 
\exp(i \beta[- \cos \mu \theta_{\textbf{k}} 
\sin\omega t + \mu \nu  \sin \mu \theta_{\textbf{k}} (\cos \omega t-1)])$.
Note that $ u \big(\theta_{\textbf{k}},t)$ is also a time-periodic function.
The quasienergy of $ \psi_{({\bf k},\eta)}^{(0)}(t)$ is exactly the same 
as the zero field case. It tells us that all the quasienergy gaps 
appear due to the interband term $\tilde{H}_{\rm inter}(t)$.

Let the solution of Eq. (\ref{A-Schro}) be of the form:
\begin{equation}
|\psi_{\eta}({\bf k}, t) \rangle = \sum \limits_{\gamma = -1}^{1} 
a_{\eta,\gamma}(t)| \psi_{\gamma}^{(0)}({\bf k}, t) \rangle, \nn
\end{equation}
we get
\begin{widetext}
\begin{eqnarray}
\partial_{t} a_{\eta,1}(t) & = & \frac{\omega}{\sqrt{2} } 
\Big[ \sqrt{2} \mu  \cos2\phi \; a_{\eta,-1}(t) 
e^{2 i v_f k t}[u^* (\theta_{\textbf{k}},t)]^2 + \sin2\phi \;  
e^{i v_f k t}u^* (\theta_{\textbf{k}},t) a_{\eta,0}(t) \Big] f(t) \\
\partial_{\bar t} a_{\eta, 0}(t) & = & \frac{\omega}{\sqrt{2}} \Big[-a_{\eta,1}(t) 
e^{-i v_f k t} u(\theta_{\textbf{k}},t) + 
a_{\eta,-1}(t) e^{i v_f k t} u^* (\theta_{\textbf{k}},t) \Big] \sin2\phi \; f(t) \\
\partial_{\bar t} a_{\eta,-1}(t) & = & - \frac{\omega}{\sqrt{2}}  
\Big[\sqrt{2} \mu \cos2\phi \; a_{\eta,1}(t)  
e^{-2 i v_f k t}[u (\theta_{\textbf{k}},t)]^2 + \sin2\phi \;  
e^{-i v_f k t} u(\theta_{\textbf{k}},t) a_{\eta,0}(t) \Big] \; f(t).
\end{eqnarray}
\end{widetext}
Taking  $(\beta \omega /\sqrt{2})y_l(\theta_{k})$ and 
$(\beta \omega /\sqrt{2})z_l(\theta_{k})$ as the discrete Fourier transform 
of the periodic functions $[u (\theta_{\textbf{k}},t)]^2 f(t)$ and 
$u (\theta_{\textbf{k}},t) f(t)$ respectively, we have
\begin{widetext}
\begin{eqnarray}
\partial_{t} a_{\eta,1}(t) & = & \frac{\beta  \omega}{2}  
\Big[\sqrt{2} \cos2\phi \; a_{\eta,-1}(t) \sum_{l=-\infty}^{\infty}y_l^*(\theta_{k}) 
e^{i(2v_fk-l\omega) t} + \sin2\phi \; a_{\eta,0}(t) 
\sum_{l=-\infty}^{\infty}z_l^*(\theta_{k}) e^{i(v_f k - l\omega) t} \Big] \label{rwa1} \\
\partial_{t} a_{\eta,0}(t ) & = & \frac{\beta \omega}{2}  
\Big[-a_{\eta,1}(t) \sum_{l=-\infty}^{\infty}z_l(\theta_{k}) 
e^{-i(v_f k - l\omega) t} +  a_{\eta,-1}(t) \sum_{l = - \infty}^{\infty}z_l^*(\theta_{k}) 
e^{i (v_f k - l\omega) t}\Big] \sin2\phi  \label{rwa2} \\ 
\partial_{t} a_{\eta,-1}(t) & = & - \frac{\beta \omega}{2}  \Big[\sqrt{2} \cos2\phi \; a_{\eta,1}(t) 
\sum_{l=-\infty}^{\infty}y_l(\theta_{k}) 
e^{-i(2v_fk-l\omega)t} + \sin2\phi \; a_{\eta,0}(t) 
\sum_{l=-\infty}^{\infty}z_l(\theta_{k}) e^{-i(v_f k - l \omega)t} 
\Big]  \label{rwa3}.
\end{eqnarray}
\end{widetext}
Here
\begin{eqnarray} \label{zl}
z_{l}(\theta_{k}) & = & \frac{1}{\sqrt 2} 
\Big[u_{l+1}^{(1)}(\theta_{k})(- \mu \sin \mu \theta_k  + i\nu  \cos \mu \theta_k) \nn \\
& - &  u_{l-1}^{(1)}(\theta_{k})
(\mu \sin \mu \theta_k + i \nu \cos \mu \theta_k) \Big] 
\end{eqnarray}
and
\begin{eqnarray} \label{yl}
y_{l}(\theta_{k}) & = & \frac{1}{\sqrt{2}} 
\Big[u_{l+1}^{(2)}(\theta_{k})(- \mu \sin\mu \theta_k  + i \nu \cos \mu \theta_k) \nn \\
& - & 
u_{l-1}^{(2)}(\theta_{k})(\mu \sin \mu \theta_k 
+ i \nu \cos \mu \theta_k) \Big],
\end{eqnarray}
where $ u_{l}^{(j)}(\theta_{k})$ with ($j=1,2)$ is defined as
\be
u_{l}^{(j)}(\theta_{k}) =
\frac{1}{T} \int_{0}^{T}[u (\theta_{\textbf{k}},t)]^j e^{-i l \omega t} dt.
\ee
The exact expressions of $ u_{l}^{(1)}(\theta_{k}) $ and 
$ u_{l}^{(2)}(\theta_{k}) $ are obtained as 
\begin{eqnarray}
u_{l}^{(1)}(\theta_{k}) & = & e^{-i \mu  \nu \beta 
\sin \mu \theta_k} J_l(2|p(\theta_k)|) 
\bigg(\frac{-p(\theta_k)}{|p(\theta_k)|}\bigg)^l \\
u_{l}^{(2)}(\theta_{k}) & = & 
e^{- i 2 \mu \nu \beta  \sin \mu \theta_k} J_l(2|q(\theta_k)|)
\bigg(\frac{-q(\theta_k)}{|q(\theta_k)|}\bigg)^l,
\end{eqnarray}
where $J_l(x)$ is the $l$-th order Bessel fuction, 
$ p(\theta_k) =  \frac{\beta}{2}(\cos \mu \theta_k - 
i \mu \nu \sin \mu \theta_k) $
and $ q(\theta_k) =  \beta(\cos \mu \theta_k - 
i \mu \nu  \sin \mu \theta_k)$.
It is not possible to solve Eqs. (\ref{rwa1}), (\ref{rwa2}), and (\ref{rwa3}) 
in closed analytical form. 
However, owing to the high frequency of radiation, standard rotating wave 
approximation (RWA) can be used to obtain closed form expressions.

There are two frequency detuning terms namely 
$\delta_1 = 2 v_f k - m \omega $ and $ \delta_2 = v_f k - m \omega $, 
due to presence of an additional dispersionless band. 
Near the resonance points, $(\delta_{1,2} \simeq 0)$, 
the momentum values $\textbf{\textit{k}}_m$ are such that
the energy difference between the bands equals $m$ multiples of
photon energy $\hbar \omega$.

For even $m$ (excluding 0), the terms 
$y_m(\theta_k)$ and $z_{m/2}(\theta_k)$ are retained in their respective 
series. But, for odd integer $m$, we see that retaining the $m$-th term 
from $y_l(\theta_k)$ series allots $v_fk/\omega$ an odd integer value. 
So within RWA, all the terms in the $z_l(\theta_k)$ series will be 
rapidly oscillating, allowing us to discard this series altogether. 
Hence, for odd $m$, we retain only $y_m(\theta_k)$. This leads to two 
distinct cases for even and odd integers, each of which produces separate 
systems of coupled differential equations for the determination of 
Floquet quasienergies.

\textbf{Case I}: For even $m$ case, Eqs. (\ref{rwa1}), (\ref{rwa2}) and (\ref{rwa3})
become  
\begin{widetext}
\begin{eqnarray}
\partial_{t} a_{\eta,1}(t) & = & \frac{\beta \omega}{2} 
\Big[\sqrt{2} \mu \cos2\phi \; a_{\eta,-1}(t) y_{m}^*(\theta_{k}) 
e^{i\delta_1 t} + \sin2\phi \; a_{\eta,0}(t) z_{m/2}^*(\theta_{k}) 
e^{i\delta_2 t} \Big] \label{rwa1a} \\
\partial_{t} a_{\eta,0}(t) & = & \frac{\beta \omega}{2} 
\Big[-a_{\eta,1}(t) z_{m/2}(\theta_{k}) e^{-i\delta_2 t} + 
a_{\eta,-1}(t) z_{m/2}^*(\theta_{k}) e^{i\delta_2 t}\Big] \; \sin2\phi  \label{rwa1b} \\
\partial_{t} a_{\eta,-1}(t) & = & - \frac{\beta \omega}{2} 
\Big[\sqrt{2} \mu \cos2\phi \; a_{\eta,1}(t) y_{m}(\theta_{k}) e^{-i \delta_1 t} + 
\sin2\phi \; a_{\eta,0}(t) z_{m/2}(\theta_{k}) e^{-i\delta_2 t}\Big]. \label{rwa1c}
\end{eqnarray}
\end{widetext}

Note that Eq. (\ref{rwa1b}) is redundant for $\alpha=0 $ case.
Equations (\ref{rwa1a}) and (\ref{rwa1c}) with $\alpha=0$ reproduce all 
the approximate analytical results for graphene provided by Zhou and Wu \cite{Wu}.

Furthermore, the above set of equations can not be solved analytically  
unless we solve it on exact resonance i.e. $\delta_1 = \delta_2 = 0$. 
On exact resonance condition, the approximate expressions of quasienergies 
for $0 < \alpha <1 $ obtained from Eqs. (\ref{rwa1a}), (\ref{rwa1b}) and (\ref{rwa1c}) 
are $\lambda_0=0$ and 

\begin{eqnarray}
\lambda_\pm=\pm\frac{\beta}{\sqrt{2}}\sqrt{\cos^2 2\phi |y_{m}(\theta_{\textbf{k}})|^2 + 
\sin^2 2\phi |z_{m/2}(\theta_{\textbf{k}})|^2}
\end{eqnarray}

From the above expression, we see that the quasienergy is proportional to root mean 
modulus squared of the coupling parameters $y_{m}(\theta_{\textbf{k}})$ and 
$z_{m/2}(\theta_{\textbf{k}})$ weighted by terms dependent on Berry phase ($\sim \cos 2\phi$) 
of the system. The sum of the weights is unity for all $\alpha$. Since the Berry phase varies 
smoothly from $\pi$ to 0 as $\alpha$ goes 0 to 1, the weight of $y_{m}(\theta_{\textbf{k}})$ 
decreases while that of $z_{m/2}(\theta_{\textbf{k}})$ increases with $\alpha$.
The quasieigenenergy for special cases like $\alpha=0$ and $\alpha=1$ 
can be obtained easily. 
For $\alpha=0$, $\lambda_{\pm} = \pm \frac{\beta}{\sqrt{2}} 
|y_{m}(\theta_{\textbf{k}})|$. This is the same result as obtained for
monolayer graphene \cite{Wu}. 
On the other hand, for dice lattice ($\alpha=1$), we get 
$\lambda_{\pm} = \pm \frac{\beta}{\sqrt{2}} 
|z_{m/2}(\theta_{\textbf{k}})|$ and $\lambda_0 = 0$. 
For the dice lattice, the quasienergy gap between $\eta=\pm 1$ at 
the resonance point is
\be
\Delta_{m}(\theta_{\bf k}) = \sqrt{2} \beta
|z_{m/2}(\theta_{\textbf{k}})|.
\ee
The magnitude of the gap in graphene and dice lattice depends on the 
effective coupling parameters
$|y_{m}(\theta_{\textbf{k}})| $ and $|z_{m/2}(\theta_{\textbf{k}})|$,
respectively. 
The behaviour of the gap in graphene and dice lattice is quite
different.

\textbf{Case II:} For odd $m$, Eqs. (\ref{rwa1}), (\ref{rwa2}) and (\ref{rwa3}) 
can be approximated as
\begin{eqnarray}
\partial_{t} a_{\eta,1}(t) & \approx & \frac{\beta \omega \mu}{\sqrt{2} } 
\cos2\phi \; a_{\eta,-1}(t) y_{m}^*(\theta_{k}) e^{i\delta_1 t} \\
\partial_{t} a_{\eta,0}(t) & \approx & 0 \\
\partial_{t} a_{\eta,-1}(t) & \approx & - \frac{\beta  \omega\mu}{\sqrt{2} } 
\cos2\phi \; a_{\eta,1}(t) y_{m}(\theta_{k}) e^{-i\delta_1 t}.
\end{eqnarray} 
Thus for odd $m$, we obtain simplified expression of quasienergy for a given 
$\alpha$:
\be \label{odd-gap}
\lambda_\pm \approx \pm \frac{\beta}{\sqrt{2}}|y_{m}(\theta_{\textbf{k}})| 
\Big(\frac{1-\alpha^2}{1+\alpha^2} \Big)
\ee
and $\lambda_0 = 0$.
Interestingly, it shows that the gap with odd values of $m$ closes in
the dice lattice, which is in sharp contrast with the graphene case.
Although Eq. (\ref{odd-gap}) shows that $\Delta_m=0 $ for $\alpha=1$, but
this is not the case. We will get a small non-zero value of $\Delta_m$ 
on taking into account the higher-order contribution from 
Eqs. (\ref{rwa1}), (\ref{rwa2}) and (\ref{rwa3}).

The quasienergy gap is essentially determined by the Berry phase and 
two coupling parameters $|y_{m}(\theta_{\textbf{k}})| $ and 
$|z_{m/2}(\theta_{\textbf{k}})|$.
The expressions of the coupling parameters can be simplified further
by setting $\theta_{\bf k} = 0$ since 
the quasienergy spectrum is isotropic for all values of $\alpha$ 
for circularly palarized light.
On substitution $\theta_{\bf k} = 0$ into Eqs. (\ref{zl}) and (\ref{yl}), we get
\begin{eqnarray}
y_l(0) & = & \frac{i \nu}{\sqrt{2}} (-1)^{l} [J_{l-1}(2\beta) - J_{l+1}(2\beta)] \\
z_l(0) & = & \frac{i\nu}{\sqrt{2}} (-1)^{l} [J_{l-1}(\beta) - J_{l+1}(\beta)].
\end{eqnarray}
Thus, the approximate forms of  quasienergies for $\alpha=0$ (for any integer $m$)
and $\alpha=1$ (for even $m$) turn out to be
\begin{eqnarray}
(\lambda_{\pm})_{\alpha=0} & = & \pm \frac{\beta }{2} |J_{m+1}(2\beta) - J_{m-1}(2\beta)| 
\label{a0}, \\
(\lambda_{\pm})_{\alpha=1} & = & \pm \frac{\beta }{2} |J_{m/2+1}(\beta) - J_{m/2-1}(\beta)| 
\label{a1}.
\end{eqnarray}
For weak field ($\beta \ll 1$), the asymptotic forms of quasienergy gaps are obtained
from the above expressions as
\begin{eqnarray}
(\Delta_m)_{\alpha=0} & \sim & \frac{\beta^m}{(m-1)!} 
\hspace{0.1cm} {\rm and} \hspace{0.1cm}
(\Delta_m)_{\alpha=1} \sim  \frac{2(\beta/2)^{m/2}}{(m/2-1)!}.
\end{eqnarray}

\end{document}